\documentclass[11pt]{article}
\usepackage{jheppub}
\usepackage{amsmath}
\usepackage{amssymb}
\usepackage{braket}
\usepackage{bbold}
\usepackage{pifont}

\newcommand{\beqn}{\begin{eqnarray}}
\newcommand{\eeqn}{\end{eqnarray}}
\newcommand{\dd}{\mathrm{d}}
\newcommand{\nn}{\nonumber}

\newcommand{\gmn}{g_{\mu\nu}}
\newcommand{\bgmn}{\bar{g}_{\mu\nu}}
\newcommand{\bfmn}{\bar{f}_{\mu\nu}}
\newcommand{\fmn}{f_{\mu\nu}}

\newcommand{\xmn}{F_{\mu\nu}}

\title{Mass eigenstates in bimetric theory with matter coupling}
\author{Angnis~Schmidt-May}
\affiliation{Department of Physics \& 
        The Oskar Klein Centre,\\
        Stockholm University, AlbaNova University Centre, 
        SE-106 91 Stockholm, Sweden}
\emailAdd{angnis.schmidt-may@fysik.su.se}

\abstract{ In this paper we study the ghost-free bimetric action extended by a recently proposed coupling to matter through a composite metric. The equations of motion for this theory are derived using a method which avoids varying the square-root matrix that appears in the matter coupling. We make an ansatz for which the metrics are proportional to each other and find that it can solve the equations provided that one parameter in the action is fixed. In this case, the proportional metrics as well as the effective metric that couples to matter solve Einstein's equations of general relativity including a matter source. Around these backgrounds we derive the quadratic action for perturbations and diagonalize it into generalized mass eigenstates. It turns out that matter only interacts with the massless spin-2 mode whose equation of motion has exactly the form of the linearized Einstein equations, while the field with Fierz-Pauli mass term is completely decoupled. Hence, bimetric theory, with one parameter fixed such that proportional solutions exist, is degenerate with general relativity up to linear order around these backgrounds. }

\begin{document} 
\maketitle
\flushbottom

\section{Introduction and summary of results}

Ghost-free bimetric theory describes nonlinear interactions between a massive and a massless spin-2 field at the classical level. It emerged from a model for nonlinear massive gravity with flat reference metric which was developed in~\cite{deRham:2010ik, deRham:2010kj}, where it was also shown that the model is ghost free in some special cases.
A full proof of absence of ghost was first given in~\cite{Hassan:2011hr}. The formulation of massive gravity in a general reference frame~\cite{Hassan:2011vm} and the absence of ghost in this version of the theory~\cite{Hassan:2011tf, Hassan:2011ea} suggested the possibility to give dynamics to the reference metric and thereby introduce the first consistent, fully dynamical bimetric theory~\cite{Hassan:2011zd, Hassan:2011ea}. This particular family of massive gravity and bimetric theories generalizes linear Fierz-Pauli theory~\cite{Fierz:1939ix} and constitutes an exception to generic models for nonlinear spin-2 interactions which contain the Boulware-Deser ghost instability~\cite{Boulware:1972zf, Boulware:1973my}. Before the ghost-free formulation was known, interacting spin-2 fields had already been studied in great detail, see~\cite{Rosen:1940zz, Isham:1971gm, Aragone:1971kh, Chamseddine:1978yu, Aragone:1979bm, Buchbinder:1999ar, Boulanger:2000rq, Damour:2002ws, ArkaniHamed:2002sp, Blas:2007ep} for some examples. For a recent review on the ghost-free theories we refer the reader to~\cite{deRham:2014zqa}.

Bimetric theory is formulated in terms of two dynamical rank-2 tensors $\gmn$ and $\fmn$ whose kinetic terms have the usual Einstein-Hilbert structure. The bimetric action furthermore contains a nonlinear interaction potential for the two metrics whose structure is constrained by requiring the absence of the Boulware-Deser ghost. Around backgrounds on which the metrics are proportional to each other, the spectrum of spin-2 perturbations is diagonalizable into mass eigenstates and it consists of a massless and a massive spin-2 field that mix with each other at the nonlinear level~\cite{Hassan:2012wr}.

An important question to address in this type of theories is how to couple the two metrics to the matter sector which, for example, could represent the Standard Model. The requirement of avoiding the Boulware-Deser ghost reduces the number of possibilities for such couplings, which should not come as a surprise since it was already the case for the interactions among the spin-2 fields. Although it is possible for the two metrics $\gmn$ and $\fmn$ to interact with two different types of matter, the ghost instability generically reappears when both metrics are coupled to the same matter source~\cite{Yamashita:2014fga, deRham:2014naa}. Some of the interesting conceptual and phenomenological issues of doubly coupled bimetric theory have been addressed in~\cite{Akrami:2013ffa, Akrami:2014lja}.
An exception to the generically inconsistent double coupling is a particular combination of $\gmn$ and $\fmn$ into an ``effective" metric $G_{\mu\nu}$ which enters the matter Lagrangian. Although the full theory does not contain the constraint that removes the Boulware-Deser ghost, it has been shown in~\cite{deRham:2014naa, deRham:2014fha} that the terms that excite the ghost are suppressed by a mass scale that is higher than the strong-coupling scale of the theory. This implies that the theory can still be treated as a consistent effective field theory with a cut-off below which the ghost is not excited.

The matter coupling of the effective metric opens up new possibilities for the phenomenology of bimetric theory. Unfortunately, a difficulty arises when one tries to compute the equations of motion for this theory because the particular form of the effective metric complicates the variation of the action with respect to the metrics $\gmn$ and $\fmn$. This is problematic because the knowledge of classical solutions to the equations of motion is indispensable for all phenomenological applications. 

\vspace{10pt}

In this work, we study bimetric theory including the matter coupling of the effective metric in more detail. Our results are summarized below.
\begin{itemize}

\item After employing a trick that allows us to remove the problematic terms in the matter coupling, the variation of the action becomes straightforward. 
Our result for the equations of motion can be used for deriving all types of classical solutions in the full bimetric theory including matter.

\item We derive the proportional background solutions, $\fmn=c^2\gmn$ with constant $c$, in the presence of matter. These solutions are Einstein metrics which, in the absence of the matter source, reduce to the known maximally symmetric backgrounds of bimetric theory in vacuum. In contrast to the pure bimetric case, their existence requires fixing one parameter of the theory. 

\item The spectrum of perturbations for the metrics around the proportional backgrounds is computed and we define mass eigenstates by comparing their equations to those of linearized general relativity in the presence of matter.
These mass eigenstates are found to be the same as in bimetric theory in vacuum and we derive the quadratic action for the massless and the massive spin-2 mode along with the corresponding linear equations. 

\item The remarkable and unexpected outcome of the analysis of perturbations is that the effective metric that couples to matter always corresponds to the massless fluctuation around proportional backgrounds, i.e.~it satisfies the linearized Einstein equations in the presence of matter. We furthermore verify that the effective metric can be considered as a nonlinear massless spin-2 field.
This result, which is independent of the remaining parameters of the theory, implies that around proportional backgrounds differences from general relativity occur only at the nonlinear level. 
\end{itemize}

\vspace{10pt}

The paper is organized as follows. In section~\ref{review} we review the structure of ghost-free bimetric theory and its coupling to matter through the effective metric. The equations of motion for this theory are obtained in section~\ref{dereom}. In section~\ref{prop} we derive the proportional background solutions and the quadratic action of perturbations around them.
Finally, our results are discussed in section~\ref{discussion}. Some technical details are provided in the appendix.

\section{Review of bimetric theory and its coupling to matter}\label{review}
In this paper we will work with the action for interacting rank-two tensors~$\gmn$ and~$\fmn$ that couple to matter through an effective metric $G_{\mu\nu}$ which is a combination of the two and will be defined below.
The full action is of the form,
\beqn\label{act}
S=S_\mathrm{bi}+S_\mathrm{m}\,,
\eeqn
where $S_\mathrm{bi}$ involves the kinetic and interaction terms for $\gmn$ and $\fmn$,
\beqn\label{sbim}
S_\mathrm{bi}=\int\dd^4 x\Big[ m_g^2\sqrt{g}~R(g) +m_f^2\sqrt{f}~R(f) - V(g,f)\Big]   \,,
\eeqn
and $S_\mathrm{m}$ describes the coupling of matter fields $\phi^a$ to the effective metric $G_{\mu\nu}$. This coupling is assumed to have a standard form as in general relativity,\footnote{For instance, the coupling for a free scalar field would be of the form $\mathcal{L}_\mathrm{m}(G,\phi)=\sqrt{G}\,G^{\mu\nu}\partial_\mu\phi\,\partial_\nu\phi$.}
\beqn\label{matc}
S_\mathrm{m}=\int\dd^4 x~\mathcal{L}_\mathrm{m}(G,\phi^a) \,.
\eeqn
The metrics in (\ref{sbim}) possess standard Einstein-Hilbert kinetic terms, multiplied by Planck masses $m_g$ and $m_f$ setting the respective interaction strengths.
In order to avoid the Boulware-Deser ghost instability that plagues generic bimetric theories, the interaction potential~$V(g,f)$ in~(\ref{sbim}) is taken to be of the form~\cite{deRham:2010kj, Hassan:2011vm},
\beqn\label{intpot}
V(g,f)=2\mu^4\sqrt{g}\sum_{n=0}^4\beta_ne_n\big(\sqrt{g^{-1}f}\big)\,.
\eeqn
Here, $\mu$ is an arbitrary mass scale, $\beta_n$ are the interaction parameters and $e_n\big(\sqrt{g^{-1}f}\big)$ denote the elementary symmetric polynomials of the square-root matrix\footnote{Starting from the above metric formulation, one can derive an explicit condition for the existence of the square-root matrix in terms of ADM variables, yielding a bound on a combination of metric components and their spatial vierbeins~\cite{us}.
This bound turns out to be equivalent to a condition under which a certain combination of vierbeins for the two metrics is symmetrizable by a local Lorentz transformation~\cite{Deffayet:2012zc}.} $\sqrt{g^{-1}f}$ defined through $\big(\sqrt{g^{-1}f}\big)^2=g^{-1}f$. The explicit expressions for the $e_n(S)$ as functions of any matrix $S$ can be obtained from the following recursion formula,
\beqn
e_n(S)=\frac{1}{n}\sum_{k=0}^{n-1}(-1)^{k+n+1}\mathrm{Tr}(S^{n-k})e_k(S)\,,\qquad e_0(S)=1\,.
\eeqn
Due to the identity $\sqrt{g}\,e_n\big(\sqrt{g^{-1}f}\big)=\sqrt{f}\,e_{4-n}\big(\sqrt{f^{-1}g}\big)$, the structure of the potential~(\ref{intpot}) is symmetric with respect to the two metrics $\gmn$ and $\fmn$. That is to say that interchanging the metrics in $S_\mathrm{bi}$ results in an action which is of the same form but with redefined parameters.

The form of the effective metric $G_{\mu\nu}$ that enters the matter coupling~(\ref{matc}) has first been proposed in~\cite{deRham:2014naa} and further studied in~\cite{us}. It reads, 
\beqn\label{effg}
G_{\mu\nu}=a^2\gmn+2ab\,g_{\mu\rho}{\big(\sqrt{g^{-1}f}\big)^\rho}_\nu+b^2\fmn\,,
\eeqn
in which $a$ and $b$ are arbitrary constants.\footnote{A generalization of the effective metric in terms of vielbeins for theories involving more than two interacting spin-2 fields has been proposed in~\cite{Noller:2014sta}. For cosmological solutions in a theory with multiple vielbeins coupled to matter, see~\cite{Tamanini:2013xia}.}
Note that due to the matrix identity $g\sqrt{g^{-1}f}=f\sqrt{f^{-1}g}$, also the structure of the effective metric is symmetric with respect to $\gmn$ and $\fmn$, in the sense that interchanging the metrics in $G_{\mu\nu}$ does not change its form but only redefines the parameters~$a$ and~$b$.

An interesting property of the above matter coupling that has already been observed in~\cite{deRham:2014naa} is that any vacuum contribution coming from the matter sector can be absorbed into the bimetric potential by rescaling the $\beta_n$ parameters. This can be seen by considering a contribution of the form $\mathcal{L}_\mathrm{m}^\mathrm{vac}=\mu^2\sqrt{G}\,\Lambda$ with constant $\Lambda$ which becomes,
\beqn\label{vacen}
\mathcal{L}_\mathrm{m}^\mathrm{vac}=\mu^2\sqrt{ G}\,\Lambda=\mu^2\Lambda a^4\sqrt{ g}\,\det\left(\mathbb{1}+\tfrac{b}{a}\sqrt{g^{-1}f}\right)=\mu^2\Lambda a^4\sqrt{ g}\,\sum_{n=0}^4\left(\tfrac{b}{a}\right)^ne_n\big(\sqrt{g^{-1}f}\big)\,.
\eeqn
These terms can be shifted into the interaction potential~(\ref{intpot}) which afterwards contains new parameters $\beta_n'=\beta_n-\frac{\Lambda a^4}{2\mu^2}\left(\tfrac{b}{a}\right)^n$. This degeneracy will allow us to be fully general when considering matter sectors without vacuum energy.

We are interested in deriving the equations of motion following from the complete action~(\ref{act}).
The variations of the bimetric part $S_\mathrm{bi}$ with respect to $\gmn$ and $\fmn$ are well-known,\footnote{Here and in the following, tensors with upper indices are the inverses of the corresponding lower-index objects. We do not use any of the metrics to raise indices.}
\begin{align}\label{oreq}
\frac{1}{\sqrt{g}}\frac{\delta S_\mathrm{bi}}{\delta g^{\mu\nu}}&=m_g^2\mathcal{G}_{\mu\nu}(g)+\mu^4\sum_{n=0}^3(-1)^n\beta_nY^{(n)}_{\mu\nu}\,,\nn\\
\frac{1}{\sqrt{f}}\frac{\delta S_\mathrm{bi}}{\delta f^{\mu\nu}}&=m_f^2\mathcal{G}_{\mu\nu}(f)+\mu^4\sum_{n=0}^3(-1)^n\beta_{4-n}\hat{Y}^{(n)}_{\mu\nu}\,,
\end{align}
where $\mathcal{G}_{\mu\nu}(g)=R_{\mu\nu}(g)-\frac{1}{2}\gmn R(g)$ denotes the Einstein tensor and the variation of the potential gives rise to the following matrix functions,
\beqn
Y^{(n)}_{\mu\nu}&=&g_{\mu\rho}\sum_{k=0}^n(-1)^ke_k\big(\sqrt{g^{-1}f}\big)\,\big(\big(\sqrt{g^{-1}f}\big)^{n-k}\big)^\rho_{~\nu}\,,\nn\\
\hat{Y}^{(n)}_{\mu\nu}&=&f_{\mu\rho}\sum_{k=0}^n(-1)^ke_k\big(\sqrt{f^{-1}g}\big)\,\big(\big(\sqrt{f^{-1}g}\big)^{n-k}\big)^\rho_{~\nu}\,.
\eeqn
The variation of the matter coupling with respect to $\gmn$ and $\fmn$ is difficult to compute due to the appearance of the square-root matrix $\sqrt{g^{-1}f}$ in the effective metric~(\ref{effg}). Varying the square-root in the potential is much simpler because its powers appear only under the trace in the elementary symmetric polynomials. In the matter coupling, however, it multiplies the stress-energy tensor of the matter fields and in order to compute the variation of $S_\mathrm{m}$ one needs to know the variation of the square root~$\sqrt{g^{-1}f}$ with respect to $\gmn$ and $\fmn$. In principle, this variation can be obtained explicitly but the resulting expressions are expected to be lengthy and difficult to handle. 

The situation is simplified when one of the parameters $a$ and $b$ in the effective metric~(\ref{effg}) is set to zero, in which case the square-root drops out of the coupling and only one of the two metrics $\gmn$ and $\fmn$ interacts with matter. This model, however, is rather restrictive and, for instance, does not allow for proportional solutions for the metrics~\cite{Hassan:2012wr}. 

The goal of this work is to overcome the aforementioned difficulties and derive the equations of motions for the more general matter coupling of $G_{\mu\nu}$ with arbitrary $a$ and $b$.

\section{Derivation of the equations of motion}\label{dereom}

In this section we derive the equations of motion following from the full action~(\ref{act}). We start by explaining the general procedure which we employ in order to simplify the variation of the matter coupling.

\subsection{An alternative set of equations}\label{equiveq}
As outlined in the previous section, bimetric theory is formulated in terms of the independent variables $\gmn$ and $\fmn$ which possess standard Einstein-Hilbert kinetic terms. The equations obtained from varying the action~(\ref{act}) with respect to these fields are,
\beqn\label{gfeq}
\left.\frac{\delta S_\mathrm{bi}}{\delta g^{\mu\nu}}\right|_{f}+\left.\frac{\delta S_\mathrm{m}}{\delta g^{\mu\nu}}\right|_{f}=0\,,\qquad
\left.\frac{\delta S_\mathrm{bi}}{\delta f^{\mu\nu}}\right|_{g}+\left.\frac{\delta S_\mathrm{m}}{\delta f^{\mu\nu}}\right|_{g}=0\,.
\eeqn
Here, $\left.\right|_g$ means that the variation is taken with $\gmn$ kept fixed.
Varying the bimetric action $S_\mathrm{bi}$ results in the known expressions given in~(\ref{oreq}). 
On the other hand, the matter coupling involves the combination,
\beqn\label{defxmn}
\xmn\equiv g_{\mu\rho}{(\sqrt{g^{-1}f})^\rho}_\nu\,,
\eeqn
which complicates the derivation of the equation of motion for the theory including matter because the variation of the square-root matrix requires a lengthy computation. 

Our strategy here will be to derive the equations without having to vary the square-root matrix.
To this end, we first rewrite the variation of the matter coupling as,
\beqn \label{rewsmg}
\left.\frac{\delta S_\mathrm{m}(g,f)}{\delta g^{\mu\nu}}\right|_{f}=\left.\frac{\delta S_\mathrm{m}(g(F,f),f)}{\delta F_{\alpha\beta}}\right|_{f}\left.\frac{F_{\alpha\beta}}{\delta g^{\mu\nu}}\right|_{f}\,,
\eeqn
where on the right-hand side $\gmn$ in $S_\mathrm{m}$ is replaced by,
\beqn\label{grepl}
\gmn(F,f)=F_{\mu\rho}f^{\rho\sigma}F_{\sigma\nu}\,,
\eeqn
and $|_{f}$ means that the variation is taken with $\fmn$ kept fixed. The last identity follows from,
\beqn \label{idff}
Ff^{-1}F=g\sqrt{g^{-1}f}\,f^{-1}g\sqrt{g^{-1}f}=g\sqrt{g^{-1}f}\,\sqrt{f^{-1}g}\,\sqrt{f^{-1}g}\,\sqrt{g^{-1}f}=g\,,
\eeqn
which we have written in matrix notation and where we have used that the inverse\footnote{The inverse of the square-root matrix can easily be derived as follows: For two matrices $S$ and $X$, the relation $S=\sqrt{X}$ implies $S^2=X$ from which it follows that $S^{-2}=X^{-1}$ which finally gives $S^{-1}=\sqrt{X^{-1}}$.}  of the square-root matrix is
 $\big(\sqrt{g^{-1}f}\,\big)^{-1}=\sqrt{f^{-1}g}$\,.
Moreover, equation~(\ref{rewsmg}) is an identity because, after the replacement, $\gmn$ appears in $S_\mathrm{m}$ only through the combination $\xmn$ in~(\ref{defxmn}). As we will see in the next subsection, the variation of the matter action~$S_\mathrm{m}(g(F,f),f)$ with respect to $\xmn$ is straightforward. The problematic term is the Jacobian $\left.\frac{F_{\alpha\beta}}{\delta g^{\mu\nu}}\right|_{f}$ whose evaluation requires varying the square-root matrix.
Of course, one can do the same for the $\fmn$ variation, use an identity similar to the one in~(\ref{idff}) and replace,
\beqn\label{frepl}
\fmn(F,g)=F_{\mu\rho}g^{\rho\sigma}F_{\sigma\nu}\,,
\eeqn
in the matter coupling.
Then the variation of $S_\mathrm{m}$ with respect to $\fmn$ may be rewritten as\,,
\beqn\label{rewsmf}
\left.\frac{\delta S_\mathrm{m}(g,f)}{\delta f^{\mu\nu}}\right|_{g}=\left.\frac{\delta S_\mathrm{m}(g,f(F,g))}{\delta F_{\alpha\beta}}\right|_{g}\left.\frac{\delta F_{\alpha\beta}}{\delta f^{\mu\nu}}\right|_{g}\,.
\eeqn
Again this is an identity because, after the above replacement, $\fmn$ appears in $S_\mathrm{m}$ only through the combination $\xmn$ in~(\ref{defxmn}). At first sight, the way in which we write the variations in equations~(\ref{rewsmg}) and (\ref{rewsmf}) may seem slightly unfamiliar, although it is straightforward to see that their validity is simply a consequence of the chain rule for taking derivatives. In order to illustrate this further, we provide a simplified example for the identities in appendix~\ref{scex}.

The full equations of motion, obtained from varying the action~(\ref{act}) with respect to the two metrics, are thus of the following form,
\beqn\label{gfeq2}
\left.\frac{\delta S_\mathrm{bi}}{\delta g^{\mu\nu}}\right|_{f}+\left.\frac{\delta S_\mathrm{m}(g(F,f),f)}{\delta F_{\alpha\beta}}\right|_{f}\left.\frac{\delta F_{\alpha\beta}}{\delta g^{\mu\nu}}\right|_{f}&=&0\,,\nn\\
\left.\frac{\delta S_\mathrm{bi}}{\delta f^{\mu\nu}}\right|_{g}+\left.\frac{\delta S_\mathrm{m}(g,f(F,g))}{\delta F_{\alpha\beta}}\right|_{g}\left.\frac{F_{\alpha\beta}}{\delta f^{\mu\nu}}\right|_{g}&=&0\,.
\eeqn
The problem is that we cannot derive the explicit expressions for these equations without knowing the variations of $F_{\alpha\beta}$ with respect to $f^{\mu\nu}$ and $g^{\mu\nu}$.  On the other hand, we now observe that it is easy to derive the expressions of the inverse Jacobians $\left.\frac{\delta f^{\mu\nu}}{\delta F_{\alpha\beta}}\right|_{g}$ and $\left.\frac{\delta g^{\mu\nu}}{\delta F_{\alpha\beta}}\right|_{f}$ from (\ref{frepl}) and~(\ref{grepl}), respectively. In particular, this calculation does not involve varying a square-root matrix. Furthermore, by definition, the inverse Jacobians satisfy,
\beqn
\left.\frac{\delta f^{\mu\nu}}{\delta F_{\alpha\beta}}\right|_{g}\left.\frac{\delta F_{\rho\sigma}}{\delta f^{\mu\nu}}\right|_{g}=\tfrac{1}{2}\left(\delta^{\alpha}_\rho\delta^{\beta}_\sigma+\delta^{\alpha}_\sigma\delta^{\beta}_\rho\right)\,,\qquad
\left.\frac{\delta g^{\mu\nu}}{\delta F_{\alpha\beta}}\right|_{f}\left.\frac{\delta F_{\rho\sigma}}{\delta g^{\mu\nu}}\right|_{f}=\tfrac{1}{2}\left(\delta^{\alpha}_\rho\delta^{\beta}_\sigma+\delta^{\alpha}_\sigma\delta^{\beta}_\rho\right)\,.
\eeqn
Contracting (\ref{gfeq2}) with $\left.\frac{\delta g^{\mu\nu}}{\delta F_{\alpha\beta}}\right|_{f}$ and $\left.\frac{\delta f^{\mu\nu}}{\delta F_{\alpha\beta}}\right|_{g}$, respectively, therefore allows us to remove the variations of the square-root matrix. The result is,
\beqn\label{neweqgf}
\left.\frac{\delta g^{\mu\nu}}{\delta F_{\alpha\beta}}\right|_{f}\left.\frac{\delta S_\mathrm{bi}}{\delta g^{\mu\nu}}\right|_{f}+\left.\frac{\delta S_\mathrm{m}(g(F,f),f)}{\delta F_{\alpha\beta}}\right|_{f}&=&0\,,\nn\\
\left.\frac{\delta f^{\mu\nu}}{\delta F_{\alpha\beta}}\right|_{g}\left.\frac{\delta S_\mathrm{bi}}{\delta f^{\mu\nu}}\right|_{g}+\left.\frac{\delta S_\mathrm{m}(g,f(F,g))}{\delta F_{\alpha\beta}}\right|_{g}&=&0\,,
\eeqn
in which all expressions can now be computed straightforwardly.

Before coming to this calculation, let us make one short remark.
The Jacobian factors $\left.\frac{\delta g^{\mu\nu}}{\delta F_{\alpha\beta}}\right|_{f}$ and $\left.\frac{\delta f^{\mu\nu}}{\delta F_{\alpha\beta}}\right|_{g}$ are computed in appendix~\ref{vari} and from the results it may not be entirely obvious that they are invertible.\footnote{Inverting the Jacobians explicitly is equivalent to computing the square-root variation. It is precisely the contraction of the equations with the Jacobian factors that significantly simplifies the expressions for the variations of the square-root matrix.} If they were not, then the equations in (\ref{neweqgf}) could in principle allow for solutions on which the variations of $F_{\alpha\beta}$ with respect to $f^{\mu\nu}$ and $g^{\mu\nu}$ become singular and which therefore do not solve the original equations. However, the fact that the functional relations in (\ref{defxmn}) can generically be inverted to give~(\ref{grepl}) and (\ref{frepl}) ensures the invertibility of the respective Jacobians.
 Hence, except for very peculiar cases in which~(\ref{defxmn}) cannot be solved for $\gmn$ or $\fmn$, the equations in~(\ref{neweqgf}) possess the same solutions as the original $\gmn$ and $\fmn$ equations and the two sets can be regarded as equivalent.
 
 As we shall see later, for the purpose of this work we do not need to worry about obtaining additional solutions because on the ansatz we make for the metrics, the Jacobians are explicitly invertible and the equations in~(\ref{neweqgf}) are definitely equivalent to the original $\gmn$ and $\fmn$  equations.

\subsection{Varying the matter coupling}

We are now going to derive the contributions from the matter coupling which appear in the equations of motion~(\ref{neweqgf}). 
The matter action~$S_\mathrm{m}$ depends on the metrics $\gmn$ and $\fmn$ (and thus also on~$\xmn$) only through the composite metric~$G_{\mu\nu}$. 
Its variations with respect to~$\xmn$ can therefore be written as,
\beqn\label{varm}
\left.\frac{\delta S_\mathrm{m}(g(F,f),f )}{\delta F_{\alpha\beta}}\right|_{f}&=&\left.\frac{\delta S_\mathrm{m}}{\delta G^{\alpha\beta}}\frac{\delta G^{\alpha\beta}(g(F,f),f)}{\delta F_{\mu\nu}}\right|_f\,,\nn\\
\left.\frac{\delta S_\mathrm{m}(g,f(F,g))}{\delta F_{\alpha\beta}}\right|_{g}&=&\left.\frac{\delta S_\mathrm{m}}{\delta G^{\alpha\beta}}\frac{\delta G^{\alpha\beta}(g,f(F,g))}{\delta F_{\mu\nu}}\right|_g\,,
\eeqn
where $G^{\mu\nu}$ with upper indices is the inverse of $G_{\mu\nu}$ and, of course, all matter fields are held fixed when varying~$S_\mathrm{m}$.
The variation of $S_\mathrm{m}$ with respect to $G^{\mu\nu}$ depends on the matter content of the theory. Since we do not make any assumptions on the matter sector here, we do not evaluate $\frac{\delta S_\mathrm{m}}{\delta G^{\mu\nu}}$ further. Its form will be exactly as in general relativity, with the usual metric~$\gmn$ replaced by the effective metric $G_{\mu\nu}$.\footnote{For some applications it may be useful to define the stress-energy tensor for the matter source with respect to the effective metric, $T_{\mu\nu}=-\frac{1}{\sqrt{G}}\frac{\delta\mathcal S_\mathrm{m}}{\delta G^{\mu\nu}}$. For instance, in the cosmological context, where one makes homogeneous and isotropic ans\"atze for the metrics, this assumes the form of a perfect fluid.} 

When $\gmn$ is replaced by (\ref{grepl}), the effective metric~(\ref{effg}) that couples to matter becomes,
\beqn\label{effmxf}
G_{\mu\nu}(g(F,f),f)=b^2\fmn+2abF_{\mu\nu}+a^2F_{\mu\rho}f^{\rho\sigma}F_{\sigma\nu}\,.
\eeqn
Similarly, replacing $\fmn$ by (\ref{frepl}) gives,
\beqn\label{effmx}
G_{\mu\nu}(g,f(F,g))=a^2\gmn+2abF_{\mu\nu}+b^2F_{\mu\rho}g^{\rho\sigma}F_{\sigma\nu}\,.
\eeqn
These expressions can now easily be varied with respect to $\xmn$.
We only present the result here; the calculations are performed in appendix~\ref{vari}. Inserting~(\ref{Gvarx2}) and~(\ref{Gvarx}) into~(\ref{varm}), the variations of the matter coupling are obtained as,
\beqn\label{matvarxg}
\frac{\delta S_\mathrm{m}}{\delta F_{\mu\nu}}&=&\left.\frac{\delta S_\mathrm{m}}{\delta G^{\alpha\beta}}\frac{\delta G^{\alpha\beta}}{\delta F_{\mu\nu}}\right|_f
=-\frac{\delta S_\mathrm{m}}{\delta G^{\alpha\beta}}\big(2ab\,\delta^\mu_\rho\delta^\nu_\sigma+a^2f^{\mu\lambda} F_{\lambda\sigma}\delta^\nu_\rho+a^2f^{\nu\lambda} F_{\lambda\sigma}\delta^\mu_\rho\big)G^{\alpha\rho}G^{\beta\sigma}
\,,\nn\\
\frac{\delta S_\mathrm{m}}{\delta F_{\mu\nu}}&=&\left.\frac{\delta S_\mathrm{m}}{\delta G^{\alpha\beta}}\frac{\delta G^{\alpha\beta}}{\delta F_{\mu\nu}}\right|_g
=-\frac{\delta S_\mathrm{m}}{\delta G^{\alpha\beta}}\big(2ab\,\delta^\mu_\rho\delta^\nu_\sigma+b^2g^{\mu\lambda} F_{\lambda\sigma}\delta^\nu_\rho+b^2g^{\nu\lambda} F_{\lambda\sigma}\delta^\mu_\rho\big)G^{\alpha\rho}G^{\beta\sigma}\,.
\eeqn
The expressions on the right-hand side become functions of $\gmn$ and $\fmn$ alone once one replaces $\xmn$ by~(\ref{defxmn}) and the effective metric $G_{\mu\nu}$ using~(\ref{effg}).

\subsection{Complete equations}\label{seccompleq}

We now combine the results found in the previous subsections to obtain the equations of motion for bimetric theory including its coupling to matter. The expressions for the Jacobian factors $\left.\frac{\delta g^{\mu\nu}}{\delta F_{\alpha\beta}}\right|_{f}$ and $\left.\frac{\delta f^{\mu\nu}}{\delta F_{\alpha\beta}}\right|_{g}$ are derived in (\ref{fvarg}) and (\ref{fvarx}).
Inserting these together with (\ref{oreq}) and (\ref{matvarxg}) into the equations of motion in~(\ref{neweqgf}), we find,
\beqn\label{cool}
0&=&\sqrt{g}\Big[m_g^2\mathcal{G}_{\rho\sigma}(g)+\mu^4\sum_{n=0}^3(-1)^n\beta_nY^{(n)}_{\rho\sigma}\Big]
(F^{\rho\mu}g^{\sigma\nu}+F^{\rho\nu}g^{\sigma\mu})\nn\\
&~&\hspace{68pt}+
\big(2ab\,\delta^\mu_\rho\delta^\nu_\sigma+a^2f^{\mu\lambda} F_{\lambda\sigma}\delta^\nu_\rho+a^2f^{\nu\lambda} F_{\lambda\sigma}\delta^\mu_\rho\big)G^{\alpha\rho}G^{\beta\sigma}\frac{\delta S_\mathrm{m}}{\delta G^{\alpha\beta}}\,,\nn\\
0&=&\sqrt{f}\Big[m_f^2\mathcal{G}_{\rho\sigma}(f)+\mu^4\sum_{n=0}^3(-1)^n\beta_{4-n}\hat{Y}^{(n)}_{\rho\sigma}\Big]
(F^{\rho\mu}f^{\sigma\nu}+F^{\rho\nu}f^{\sigma\mu})\nn\\
&~&\hspace{73pt}+\big(2ab\,\delta^\mu_\rho\delta^\nu_\sigma+b^2g^{\nu\lambda} F_{\lambda\sigma}\delta^\mu_\rho+b^2g^{\mu\lambda} F_{\lambda\sigma}\delta^\nu_\rho\big)G^{\alpha\rho}G^{\beta\sigma}\frac{\delta S_\mathrm{m}}{\delta G^{\alpha\beta}}\,.
\eeqn 
As usual, $F^{\mu\nu}$ with upper indices denotes the inverse of $\xmn$.
The final result can easily be expressed in terms of the original variables by making the replacements $\xmn=g_{\mu\rho}{(\sqrt{g^{-1}f})^\rho}_\nu$ and $G_{\mu\nu}=a^2\gmn+2abg_{\mu\rho}{(\sqrt{g^{-1}f})^\rho}_\nu+b^2\fmn$.
Alternatively, one could also regard $\fmn$ and~$G_{\mu\nu}$ as functions of $\gmn$ and $\xmn$, given through~(\ref{frepl}) and~(\ref{effmx}), respectively. In this case, when solving the equations, one would make ans\"atze for $\gmn$ and $\xmn$ instead of the original metrics.
While this choice has the advantage of avoiding the square-root matrix, the first option may be preferable because it simplifies the structure of the kinetic terms. For ans\"atze that allow a straightforward evaluation of the square-root matrix (e.g.~for diagonal metrics), expressing the equations in terms of $\gmn$ and $\fmn$ is definitely the better strategy.

Taking $a=1$ and $b=0$, one recovers the well-known bimetric equations with only $\gmn$ coupled to matter. 
This can be verified straightforwardly by observing that the matter coupling drops out of the second equation in (\ref{cool}), which then becomes the bimetric $\fmn$ equation contracted with the (invertible) operator $(F^{\rho\mu}f^{\sigma\nu}+F^{\rho\nu}f^{\sigma\mu})$. 
Moreover, in this case we have $G_{\mu\nu}=\gmn$ and the first equation in (\ref{cool}) becomes the bimetric $\gmn$ equation including matter and contracted with $(F^{\rho\mu}g^{\sigma\nu}+F^{\rho\nu}g^{\sigma\mu})$. Of course, in a similar manner, one can also rearrive at the bimetric equations with only $\fmn$ coupled to matter by setting $a=0$ and $b=1$.

\section{Proportional backgrounds and their perturbations}\label{prop}

Proportional background solutions and the spectrum of fluctuation around them for bimetric theory in vacuum have already been studied in detail~\cite{Hassan:2012wr}. We will briefly review the results of this analysis before proceeding to the full theory including the matter coupling.

\subsection{Bimetric theory in vacuum}

In the absence of matter, the equations of motion for bimetric theory are given by the vanishing of~(\ref{oreq}). In order to find proportional solutions for the two metrics, we make an ansatz $\bfmn=c^2\bgmn$ with arbitrary constant $c$ to be determined by the equations. On this ansatz, the matrix functions $Y^{(n)}_{\mu\nu}$ and $\hat{Y}^{(n)}_{\mu\nu}$ in~(\ref{oreq}) that are obtained from varying the interaction potential become proportional to the metric $\bgmn$ since $g^{-1}f$ becomes proportional to the identity matrix. As a consequence, the bimetric equations reduce to two copies of Einstein's equations for the metric $\bgmn$,
\beqn\label{probbge}
\mathcal{G}_{\mu\nu}(\bar{g})+\Lambda_g\bgmn=0\,,\qquad
\mathcal{G}_{\mu\nu}(\bar{g})+\Lambda_f\bgmn=0\,,
\eeqn
where the cosmological constants are functions of the proportionality constant $c$ as well as the parameters in the bimetric action,
\beqn
\Lambda_g=\frac{\mu^4}{m_g^2}(\beta_0+3c\beta_1+3c^2\beta_2+c^3\beta_3)\,,\quad
\Lambda_f=\frac{\mu^4}{c^2m_f^2}(c\beta_1+3c^2\beta_2+3c^3\beta_3+c^4\beta_4)\,.
\eeqn
Consistency of the two equations in~(\ref{probbge}) requires $\Lambda_g=\Lambda_f$, which determines~$c$ in terms of the bimetric parameters.\footnote{An exception to this is the special parameter choice $\alpha^4\beta_0=3\alpha^2\beta_2=\beta_4$, $\beta_1=\beta_3=0$, which has been suggested as a model for nonlinear partial masslessness~\cite{Hassan:2012gz, Hassan:2012rq, Hassan:2013pca}.} The solutions for both $\gmn$ and $\fmn$ are maximally symmetric Einstein metrics with cosmological constant $\Lambda_g$. 

Equivalently, we could have derived the above solutions from~(\ref{cool}) in which the matter source is set to zero. On the proportional ansatz, the Jacobians in~(\ref{neweqgf}) both become proportional to $(\bar{g}^{\rho\mu}\bar{g}^{\sigma\nu}+\bar{g}^{\rho\nu}\bar{g}^{\sigma\mu})$. The equations can then be contracted with $\bar{g}_{\mu\alpha}\bar{g}_{\nu\beta}$ which reduces them to (\ref{probbge}). This explicitly verifies that, for the proportional ansatz, no additional solutions are introduced by solving the new instead of the original equations.

Depending on the choice of $\beta_n$ parameters, non-proportional maximally symmetric solutions may also exist~\cite{Gratia:2013uza, Hassan:2014vja}, but the proportional backgrounds are the only solutions that allow the fluctuations of the metrics to be diagonalized into spin-2 mass eigenstates~\cite{Hassan:2012wr}. 
The spectrum around proportional backgrounds consists of one massless and one massive perturbation that are linear superpositions of the metric fluctuations $\delta\gmn$ and $\delta\fmn$,
\beqn\label{flucdef}
\delta G_{\mu\nu}={N_G}\left(\delta\gmn+\alpha^2\delta\fmn\right)\,,\qquad
\delta M_{\mu\nu}={N_M}\left(\delta\fmn-c^2\delta\gmn\right)\,,
\eeqn
where $N_G$ and $N_M$ are constants that may be fixed to canonically normalize the kinetic terms and $\alpha= m_f/m_g$.
The equations for these fluctuations are obtained from linearizing~(\ref{oreq}) in $\delta\gmn$ and $\delta\fmn$ and building linear combinations of the equations. Alternatively, one can derive them from~(\ref{cool}) in which the matter source is set to zero. 
The result reads~\cite{Hassan:2012wr},
\begin{subequations}
\label{fluceq}
\begin{align}
\mathcal{E}^{\rho\sigma}_{\mu\nu}\delta G_{\rho\sigma}-\Lambda_g\left(\delta G_{\mu\nu}-\tfrac{1}{2}\bar{g}_{\mu\nu}\bar{g}^{\rho\sigma}\delta G_{\rho\sigma}\right)=&~0\,,\label{Geq}\\
\mathcal{E}^{\rho\sigma}_{\mu\nu}\delta M_{\rho\sigma}-\Lambda_g\left(\delta M_{\mu\nu}-\tfrac{1}{2}\bar{g}_{\mu\nu}\bar{g}^{\rho\sigma}\delta M_{\rho\sigma}\right)+\tfrac{m_\mathrm{FP}^2}{2}\left(\delta M_{\mu\nu}-\bar{g}_{\mu\nu}\bar{g}^{\rho\sigma}\delta M_{\rho\sigma}\right)=&~0\,,\label{Meq}
\end{align}
\end{subequations}
where the kinetic structure is given by the linearized Einstein operator,
\begin{align}\label{kinop}
\mathcal{E}^{\rho\sigma}_{\mu\nu}\delta G_{\rho\sigma}=-\tfrac{1}{2}\big(
\delta^\rho_\mu\delta^\sigma_\nu\bar{\nabla}^2+\bar{g}^{\rho\sigma}\bar{\nabla}_\mu\bar{\nabla}_\nu&-\delta^\rho_\mu\bar{\nabla}^\sigma\bar{\nabla}_\nu-\delta^\rho_\nu\bar{\nabla}^\sigma\bar{\nabla}_\mu\nn\\
&-\bar{g}_{\mu\nu}\bar{g}^{\rho\sigma}\bar{\nabla}^2+\bar{g}_{\mu\nu}\bar{\nabla}^\rho\bar{\nabla}^\sigma
\big) \delta G_{\rho\sigma}\,.
\end{align}
and the Fierz-Pauli mass of the massive fluctuation is,
\beqn\label{fpmass}
m^2_\mathrm{FP}\equiv \frac{\mu^4}{m_g^2}\left(1+\alpha^{-2}c^{-2}\right)\left(c\beta_1+2c^2\beta_2+c^3\beta_3\right)\,,\qquad
\alpha\equiv m_f/m_g\,.
\eeqn
It is worth emphasizing that the proportional background solutions for bimetric theory in vacuum exist for general values of the parameters in the bimetric action.\footnote{This holds provided that the condition $\Lambda_g=\Lambda_f$ allows for a valid solution for $c$. Some solutions for $c$ can be problematic, for example, if they lead to a value for the Fierz-Pauli mass that satisfies $m^2_\mathrm{FP}<\frac{2}{3}\Lambda_g$, violating the Higuchi bound~\cite{Higuchi:1986py}. } The situation will change when matter is introduced.

\subsection{Including matter}\label{matprop}

Next we would like to invoke the equations of motion obtained in the previous section to re-derive the proportional background solutions and their perturbation spectrum for bimetric theory including the new coupling of the effective metric $G_{\mu\nu}$ to matter. Before doing so, we briefly comment on the notion of mass in the presence of matter sources.

\subsubsection{Mass eigenstates in the presence of matter}\label{massnot}

Clearly, the bimetric equations including matter sources will no longer admit backgrounds around which the notion of mass is well-defined, i.e.~those that solve Einstein's equations in vacuum. Nevertheless, here we will introduce a notion of spin-2 mass as an analogy to general relativity.
In general, such a definition can only make sense for backgrounds around which the spectrum of fluctuations can be diagonalized in terms of spin-2 fields (i.e.~without breaking covariance). As in bimetric theory in vacuum, these are the proportional backgrounds which happen to coincide with solutions to Einstein's equations including matter sources.
Around such backgrounds we define the notion of mass as follows: We call a perturbation ``massless" if it satisfies the linearized Einstein equations of general relativity in the presence of matter. If its equation in addition contains a Fierz-Pauli mass term, the perturbation is referred to as ``massive". According to this definition, a linear equation for the massless spin-2 fluctuation is degenerate with linearized general relativity whereas the equation for the massive field is the same as in linearized massive gravity.

Below we will encounter one important difference to bimetric theory in vacuum: Proportional backgrounds (and hence backgrounds for which our notion of mass makes sense) do not exist for all parameter values. This means that for general choices of parameters in the bimetric action, the notion of spin-2 mass eigenstates does not exist. On the other hand, since proportional backgrounds are degenerate with solutions to Einstein's equations, they are favoured by observations which confirm general relativity to a high level of precision. Fixing the parameters such that the theory admits solutions which are compatible with data is physically well-motivated and hence requiring the existence of proportional backgrounds is a good starting point for studying differences between bimetric theory and general relativity in perturbation theory. Our aim here is to show that, once this parameter choice has been made to ensure the compatibility of bimetric theory with observations at the background level, also the linear fluctuation equations will be the same as in general relativity because only the ``massless" mode interacts with matter.

\subsubsection{Background}\label{matprop1}
As before, we make the ansatz $\bfmn=c^2\bgmn$ with arbitrary constant $c$ to be determined by the equations.\footnote{In the absence of matter, a conformal ansatz $\bfmn=c(x)^2\bgmn$ reduces to the $c=\,$const.~case due to the Bianchi constraint. When the effective metric couples to matter, this is no longer obvious and here we restrict ourselves to constant $c$ in the ansatz.} On this ansatz, we have,
\beqn
\xmn=\bar{g}_{\mu\rho}{\Big(\sqrt{\bar{g}^{-1}\bar{f}}\,\Big)^\rho}_\nu=c\,\bgmn\,,
\eeqn
and the effective metric reduces to,
\beqn
\bar{G}_{\mu\nu}=a^2\bgmn+2ab\,\bar{g}_{\mu\rho}{\Big(\sqrt{\bar{g}^{-1}\bar{f}}\,\Big)^\rho}_\nu+b^2\bfmn=(a+bc)^2\bgmn\,.
\eeqn
Note that we have to demand $a\neq- bc$ in order to ensure $\bar{G}_{\mu\nu}\neq 0$.
As already discussed for bimetric theory in vacuum, the Jacobian factors in~(\ref{neweqgf}) become manifestly invertible on the proportional ansatz.
Hence, deriving the proportional solutions from~(\ref{cool}) is equivalent to obtaining them from the original $\gmn$ and $\fmn$ equations. 
Evaluated on the ansatz, the equations of motion in~(\ref{cool}) become,
\beqn\label{nbgeqc}
0&=&2c^{-1}m_g^2\Big[\mathcal{G}_{\mu\nu}(\bar{g})+\Lambda_g\bgmn\Big]
+
2(ab+a^2c^{-1})\left.\frac{1}{\sqrt{\bar{G}}}\frac{\delta\mathcal{L}_\mathrm{m}}{\delta G^{\mu\nu}}\right|_{G=\bar{G}}\,,\nn\\
0&=&2c\alpha^2m_g^2\Big[\mathcal{G}_{\mu\nu}(\bar{g})+\Lambda_f\bgmn\Big]
+
2(ab+b^2c)\left.\frac{1}{\sqrt{\bar{G}}}\frac{\delta\mathcal{L}_\mathrm{m}}{\delta G^{\mu\nu}}\right|_{G=\bar{G}}\,,
\eeqn
where again $\alpha= m_f/m_g$. 
Since any vacuum energy contribution coming from $\mathcal{L}_\mathrm{m}$ can be absorbed into the interaction parameters of the bimetric potential (see equation~(\ref{vacen})) we can assume without loss of generality that $\mathcal{L}_\mathrm{m}$ contains no vacuum energy. Then, consistency among the above equations requires that the relative factors between the curvature, the vacuum and the matter contributions are the same in both sets of equations. This means that for the existence of proportional backgrounds we must have,
\beqn
\frac{c^{-1}}{ab+a^2c^{-1}}=\frac{c\alpha^2}{ab+b^2c}
\qquad
\Lambda_g=\Lambda_f\,.
\eeqn
The first of these can be simplified such that we arrive at,
\beqn\label{conscond2}
\frac{b}{a}=c\alpha^2
\qquad
\Lambda_g=\Lambda_f\,.
\eeqn
Our ansatz only contains one free parameter $c$ whose value will be determined by one of the two conditions. The other condition will in general not be satisfied by the theory. This means that for general parameters in the action proportional background solutions do not exist; their existence requires fixing one parameter of the theory. Since $\Lambda_g=\Lambda_f$ is the condition that determines $c$ in the absence of matter, it is most intuitive to think of this condition as determining $c$ also in this case. The solutions for the proportionality constant of the backgrounds are thus the same as in vacuum and $c$ becomes a function of the $\beta_n$ parameters, corresponding to the roots of the fourth-order polynomial equation $\Lambda_g=\Lambda_f$. Now the second equation, $\frac{b}{a}=\alpha^{2}c$, requires fixing one of the parameters $a$, $b$,~$m_g$ or~$m_f$.\footnote{As in the case without matter sources, an exception is the particular parameter choice $\alpha^4\beta_0=3\alpha^2\beta_2=\beta_4$, $\beta_1=\beta_3=0$, for which the equation $\Lambda_g=\Lambda_f$ is automatically satisfied for any value of $c$. In this case, none of the parameters $a$, $b$, $m_g$ or $m_f$ need to be fixed, but now the condition $\frac{b}{a}=\alpha^{2}c$ instead determines $c$.} As a consequence, the effective metric will depend on the $\beta_n$ in a rather complicated way. This means that the relative strength with which the two metrics $\gmn$ and $\fmn$ couple to matter is no longer a free parameter but gets related to a combination of interaction parameters.

Observe that the above results confirm the well-known fact that proportional solutions do not exist when only one of the two metrics, $\gmn$ or $\fmn$, is coupled to matter. In this case we have $a=0$ or $b=0$, respectively, and the condition $\frac{b}{a}=\alpha^{2}c$ cannot be satisfied.

Once the conditions~(\ref{conscond2}) are met, the equations in~(\ref{nbgeqc}) reduce to two copies of the same Einstein equation for the metric $\bar{G}_{\mu\nu}=a^2(1+\alpha^2c^2)^2\bgmn$ which reads,
\beqn\label{bgeqg}
\mathcal{G}_{\mu\nu}(\bar{G})+\Lambda_G\bar{G}_{\mu\nu}
=\frac{1}{M_\mathrm{P}^2}\bar{T}_{\mu\nu},
\eeqn
where we have used,
\beqn\label{lambdampl}
\Lambda_G= \frac{\Lambda_g}{a^{2}(1+\alpha^2c^2)^{2}}\,,\qquad
M_\mathrm{P}^2=\frac{m_g^{2}}{a^2(1+\alpha^2c^2)}\,,\qquad
\bar{T}_{\mu\nu}=-\frac{1}{\sqrt{\bar{G}}}\left.\frac{\delta\mathcal{L}_\mathrm{m}}{\delta G^{\mu\nu}}\right|_{G=\bar{G}}\,.
\eeqn
This shows that proportional backgrounds correspond to Einstein solutions for the metric $\bar{G}_{\mu\nu}$ with cosmological constant~$\Lambda_G$ and effective Planck mass $M_\mathrm{P}$, which are functions of the parameters in the action. The situation is very similar to bimetric theory in vacuum, except that the existence of the proportional backgrounds now requires fixing one of the parameter combinations $b/a$ or $m_f/m_g$.

\subsubsection{Linear perturbations}

As a next step, we consider the perturbation equations around the proportional backgrounds. Let $c$ be determined by the condition $\Lambda_g=\Lambda_f$ and let us assume that the condition $\frac{b}{a}=\alpha^{2}c$ is satisfied by fixing one of the parameters, say $b$, such that the proportional background solutions with $\bfmn=c^2\bgmn$ exist. The nonlinear expression for the effective metric becomes,
\beqn
G_{\mu\nu}=a^2\Big(\gmn+2\alpha^2 c \,g_{\mu\rho}{\big(\sqrt{g^{-1}f}\big)^\rho}_\nu+\alpha^4c^2\fmn\Big)\,.
\eeqn
Its perturbations around the proportional backgrounds can straightforwardly be computed, using $\delta \sqrt{g^{-1}f}=\frac{1}{2c}(\delta f-c^2\delta g)$,
\beqn
\delta G_{\mu\nu}
&=&a^2\Big(\delta\gmn+2\alpha^2c^2\,\delta\gmn+\alpha^2\left(\delta\fmn-c^{2}\delta\gmn\right)+\alpha^4c^2\delta\fmn\Big)\nn\\
&=&a^2(1+\alpha^2c^2)\Big(\delta\gmn+\alpha^2\delta\fmn\Big)\,.
\eeqn
Remarkably, this is exactly the massless fluctuation~(\ref{flucdef}) of bimetric theory in vacuum with the normalization $N_G=a^2(1+\alpha^2c^2)$, which means that the effective metric that couples to matter is massless around the proportional solutions according to our definition in section~\ref{massnot}. Since these are the only backgrounds that admit a clear definition of mass, the perturbations of the field $G_{\mu\nu}$ are massless whenever a notion of mass exists.

We now compute the full quadratic action for perturbations around proportional backgrounds. First, we observe that at the quadratic level the matter coupling takes the form,
\beqn\label{linmc}
S^{(2)}_\mathrm{m}=\int\dd^4x~\delta( G^{\mu\nu})~\delta\left(\frac{\delta\mathcal{L}_\mathrm{m}}{\delta G^{\mu\nu}}\right)=\int\dd^4x\,\sqrt{\bar{G}}~\bar{G}^{\mu\rho}\bar{G}^{\nu\sigma}\delta G_{\rho\sigma}\delta 
T_{\mu\nu}\,,
\eeqn
where $\delta T_{\mu\nu}=-\frac{1}{\sqrt{\bar{G}}}\,\delta\left(\frac{\delta\mathcal{L}_\mathrm{m}}{\delta G^{\mu\nu}}\right)$ and $\delta\left(\frac{\delta\mathcal{L}_\mathrm{m}}{\delta G^{\mu\nu}}\right)$ is the linearization of $\frac{\delta\mathcal{L}_\mathrm{m}}{\delta G^{\mu\nu}}$.
In order to derive the quadratic action for the fluctuations on the bimetric side, we first note that (\ref{flucdef}) may be reversed to give the perturbations of the original metrics in terms of the mass eigenstates,
\beqn
\delta\gmn=\tfrac{1}{1+\alpha^2c^2}\left(\tfrac{1}{N_G}\delta G_{\mu\nu}-\tfrac{\alpha^2}{N_M}\delta M_{\mu\nu}\right)
\,,\quad
\delta\fmn=\tfrac{1}{1+\alpha^2c^2}\left(\tfrac{c^2}{N_G}\delta G_{\mu\nu}+\tfrac{1}{N_M}\delta M_{\mu\nu}\right)\,.
\eeqn
We use these expressions in the bimetric part of the quadratic action which can be most easily computed in terms of $\delta\gmn$ and $\delta\fmn$.
After also including the above expression for the matter coupling, we obtain,
\begin{align}
S^{(2)}&=~-\tfrac{m_g^2}{N_G^2(1+\alpha^2c^2)}\int\dd^4x\,\sqrt{\bar{g}}~\Big(
\delta G_{\mu\nu}\bar{\mathcal{E}}^{\mu\nu\rho\sigma}\delta G_{\rho\sigma}+\Lambda_g\,\delta G_{\mu\nu}\bar{g}^{\mu\rho}\bar{g}^{\nu\sigma}\delta G_{\rho\sigma}\Big)\nn\\
&~~~~~~~~~~~~~~~~~~~+\int\dd^4x\,\sqrt{\bar{G}}~ \bar{G}^{\mu\rho}\bar{G}^{\nu\sigma}\delta G_{\rho\sigma}\delta T_{\mu\nu}\nn\\
&~~~~-\tfrac{c^{-2}m_f^2}{N_M^2(1+\alpha^2c^2)}\int\dd^4x\,\sqrt{\bar{g}}~\Big(\delta M_{\mu\nu}\bar{\mathcal{E}}^{\mu\nu\rho\sigma}\delta M_{\rho\sigma}+\Lambda_g\,\delta M_{\mu\nu}\bar{g}^{\mu\rho}\bar{g}^{\nu\sigma}\delta M_{\rho\sigma}\nn\\
&\hspace{170pt}+\tfrac{{m}_\mathrm{FP}^2}{2}\delta M_{\mu\nu}\left(\bar{g}^{\mu\rho}\bar{g}^{\nu\sigma}-\bar{g}^{\mu\nu}\bar{g}^{\rho\sigma}\delta M_{\rho\sigma}\right)\Big)\,,
\end{align}
where now the linearized Einstein operator reads,
\begin{align}\label{kinop2}
\bar{\mathcal{E}}^{\rho\sigma}_{\mu\nu}\delta G_{\rho\sigma}=-\tfrac{1}{2}\big(
\delta^\rho_\mu\delta^\sigma_\nu\bar{\nabla}^2&+\bar{g}^{\rho\sigma}\bar{\nabla}_\mu\bar{\nabla}_\nu-\delta^\rho_\mu\bar{\nabla}^\sigma\bar{\nabla}_\nu-\delta^\rho_\nu\bar{\nabla}^\sigma\bar{\nabla}_\mu\nn\\
&-\bar{g}_{\mu\nu}\bar{g}^{\rho\sigma}\bar{\nabla}^2+\bar{g}_{\mu\nu}\bar{\nabla}^\rho\bar{\nabla}^\sigma
-\bar{g}_{\mu\nu}\bar{R}^{\rho\sigma}+\delta^\rho_\mu\delta^\sigma_\nu\bar{R}
\big) \delta G_{\rho\sigma}\,,
\end{align}
in which $\bar{R}_{\mu\nu}$ is the background curvature of $\bgmn$.
Note that the bimetric part of the quadratic action is written with respect to the background metric $\bgmn$, while the matter coupling is more naturally expressed in terms of~$\bar{G}_{\mu\nu}$ which differs from $\bgmn$ by a constant scaling. In order to formulate the whole action with respect to the same background metric~$\bar{G}_{\mu\nu}$ that solves~(\ref{bgeqg}), we replace,
\beqn
\bar{g}^{\mu\nu}&=&a^{2}(1+{\alpha^2c^2})^2\bar{G}^{\mu\nu}\,, \qquad 
\,\sqrt{\bar{g}}~=~a^{-4}(1+{\alpha^2c^2})^{-4}\sqrt{\bar{G}}\,,\nn\\
N_G&=&a^2(1+\alpha^2c^2)\,,\qquad 
\,~~~~~~~~\Lambda_g=a^{2}(1+{\alpha^2c^2})^{2}\Lambda_G\,,\nn\\ 
m_\mathrm{FP}^2&=&a^{2}(1+{\alpha^2c^2})^{2}\bar{m}_\mathrm{FP}^2\,,\qquad
m_g^2=a^{2}(1+\alpha^2c^2)M_\mathrm{P}^2\,.
\eeqn
Moreover, we choose to fix the normalization of the massive mode to $N_M=c^{-1}\alpha N_G$, such that we finally get,
\begin{align}\label{quadact}
S^{(2)}&=~-M_\mathrm{P}^2\int\dd^4x\,\sqrt{\bar{G}}~\Big(
\delta G_{\mu\nu}\tilde{\mathcal{E}}^{\mu\nu\rho\sigma}\delta G_{\rho\sigma}+\Lambda_G\,\delta G_{\mu\nu}\bar{G}^{\mu\rho}\bar{G}^{\nu\sigma}\delta G_{\rho\sigma}\Big)\nn\\
&~~~~~~~~~~+\int\dd^4x\,\sqrt{\bar{G}}~ \bar{G}^{\mu\rho}\bar{G}^{\nu\sigma}\delta G_{\rho\sigma}\delta T_{\mu\nu}\nn\\
&~~~~\,-M_\mathrm{P}^2\int\dd^4x\,\sqrt{\bar{G}}~\Big(\delta M_{\mu\nu}{\tilde{\mathcal{E}}}^{\mu\nu\rho\sigma}\delta M_{\rho\sigma}+\Lambda_G\,\delta M_{\mu\nu}\bar{G}^{\mu\rho}\bar{G}^{\nu\sigma}\delta M_{\rho\sigma}\nn\\
&\hspace{185pt}+\tfrac{\bar{m}_\mathrm{FP}^2}{2}\delta M_{\mu\nu}\left(\bar{G}^{\mu\rho}\bar{G}^{\nu\sigma}-\bar{G}^{\mu\nu}\bar{G}^{\rho\sigma}\delta M_{\rho\sigma}\right)\Big)\,,
\end{align}
which is the quadratic action for a massless spin-2 field coupled to matter and a decoupled massive Fierz-Pauli field. It is now formulated with respect to the background metric $\bar{G}_{\mu\nu}$. In particular, $\tilde{\mathcal{E}}^{\rho\sigma}_{\mu\nu}$ is of the same form as in (\ref{kinop2}) but with $\bgmn$ replaced by $\bar{G}_{\mu\nu}$, $\Lambda_G$ is the cosmological constant and $M_\mathrm{P}$ is the effective Planck mass  for $\bar{G}_{\mu\nu}$ as defined in~(\ref{lambdampl}). In the new background metric, $\bar{m}_\mathrm{FP}$ is the Fierz-Pauli mass of the massive spin-2 mode.
Varying the action with respect to the massless fluctuation $\delta G_{\mu\nu}$ and the massive fluctuation $\delta M_{\mu\nu}$, respectively, gives the following equations,
\begin{align}
&\tilde{\mathcal{E}}^{\rho\sigma}_{\mu\nu}\delta G_{\rho\sigma}+\Lambda_G\delta G_{\mu\nu}=\frac{1}{2M_{\mathrm{P}}^2}\delta T_{\mu\nu}\,,\nn\\
&\tilde{\mathcal{E}}^{\rho\sigma}_{\mu\nu}\delta M_{\rho\sigma}+\Lambda_G\delta M_{\mu\nu}+\tfrac{\bar{m}_\mathrm{FP}^2}{2}\left(\delta M_{\mu\nu}-\bar{G}_{\mu\nu}\bar{G}^{\rho\sigma}\delta M_{\rho\sigma}\right)=0\,,
\end{align}
in which we have lowered the indices with $\bar{G}_{\mu\nu}$.
We conclude that matter only interacts with the massless spin-2 field at the linear level, while the massive field is completely decoupled. This shows that not only the proportional backgrounds but also the linear theory around them is degenerate with general relativity.

\subsection{The nonlinear massless field}\label{massless}

In~\cite{Hassan:2012wr} an attempt was made to rewrite bimetric theory in terms of combinations of~$\gmn$ and~$\fmn$ which can be interpreted as nonlinear mass eigenstates. The same reference also provided conditions for classifying a nonlinear combination of $\gmn$ and $\fmn$ as ``massless". 
Since here we have shown that the fluctuations of the effective metric $G_{\mu\nu}$ that couples to matter are massless, it now becomes interesting to check whether $G_{\mu\nu}$ also satisfies the criteria for being a nonlinear massless field. 
To this end, let us write $G_{\mu\nu}$ as,
\beqn
G_{\mu\nu}=g_{\mu\rho}{\Phi^{\rho}}_{\nu}\,, \qquad
{\Phi^{\rho}}_{\nu}=a^2\Big(\delta^{\rho}_{\nu}+2\alpha^2c{(\sqrt{g^{-1}f})^{\rho}}_{\nu}+\alpha^4c^2 g^{\rho\sigma}f_{\sigma\nu}\Big)\,,
\eeqn
where, to ensure the existence of proportional backgrounds, we have fixed $b=a\alpha^2 c$. Its fluctuations around these backgrounds then are of the form,
\beqn
\delta G_{\mu\nu}=N_G(\delta\gmn+\alpha^2\delta\fmn)\,,\qquad
N_G=a^2(1+\alpha^2c^2)\,.
\eeqn
The criteria that a nonlinear massless field needs to satisfy now read~\cite{Hassan:2012wr}\,,
\beqn\label{criteria}
N_G^{-1}\bar{\Phi}^{\mu}_{~\nu}=(1+\alpha^2c^2)\delta^{\mu}_{\nu}\,,\qquad
N_G^{-1}\left.\frac{\delta {\Phi}^{\mu}_{~\nu}}{\delta{(\sqrt{g^{-1}f})^\rho}_\sigma}\right|_{f=c^2g}=2c\alpha^2\delta^\mu_\rho\delta_\nu^\sigma\,,
\eeqn
where $\bar{\Phi}^{\mu}_{~\nu}$ denotes the background value of ${\Phi}^{\mu}_{~\nu}$ on the proportional solutions.
For our $G_{\mu\nu}$ this is of the form,
\beqn
\bar{\Phi}^{\mu}_{~\nu}=a^2(1+2\alpha^2c+\alpha^4c^4)\delta^{\mu}_{\nu}=a^2(1+\alpha^2c^2)^2\delta^{\mu}_{\nu}\,.
\eeqn
Dividing this result by $N_G$, we find that the first criterion in~(\ref{criteria}) is met. For the derivative with respect to the square-root matrix, we obtain the following expression,
\beqn
\frac{\delta {\Phi}^{\mu}_{~\nu}}{\delta{(\sqrt{g^{-1}f})^\rho}_\sigma}=a^2\Big(2\alpha^2c\delta^{\mu}_{\rho}\delta_{\nu}^{\sigma}+\alpha^4c^2 {(\sqrt{g^{-1}f})^{\mu}}_{\rho}\delta_{\nu}^{\sigma}+\alpha^4c^2 {(\sqrt{g^{-1}f})^{\sigma}}_{\nu}\delta^{\mu}_{\rho}\Big)\,.
\eeqn
On the proportional background, this reduces to,
\beqn
\left.\frac{\delta {\Phi}^{\mu}_{~\nu}}{\delta{(\sqrt{g^{-1}f})^\rho}_\sigma}\right|_{f=c^2g}=2c\alpha^2 a^2(1+\alpha^2c^2 )\delta^{\mu}_{\rho}\delta_{\nu}^{\sigma}\,,
\eeqn
and division by $N_G$ verifies that also the second criterion is satisfied. We conclude that the effective metric $G_{\mu\nu}$ is a nonlinear massless field according to the classification in~\cite{Hassan:2012wr}. 

In the remainder of this section we outline the procedure of reformulating the theory in terms of the nonlinear massless field.
As a first observation, we note that the structure of the bimetric interaction potential~(\ref{intpot}) does not change when it is rewritten in terms of $G_{\mu\nu}$ and, for instance, $\gmn$. To see this, first use the definition of the effective metric to express $\sqrt{g^{-1}f}$ in terms of $G_{\mu\nu}$ and~$\gmn$. The result is, in matrix notation,
\beqn\label{sidmg}
\sqrt{g^{-1}f}=\frac{a}{b}\Big(a^{-1}\sqrt{g^{-1}G}-\mathbb{1}\Big)\,.
\eeqn
Then we make use of the following identity of the elementary symmetric polynomials which holds for any matrix $S$,
\beqn
e_n(S-\mathbb{1})=\sum_{k=0}^n(-1)^{n-k}{4-k \choose n-k}e_k(S)\,,
\eeqn
to replace $\sqrt{g^{-1}f}$ in the bimetric interaction potential $V(g,f)$. The result is,
\beqn
V(g,G)
=2\mu^4\sqrt{g}~\sum_{n=0}^4\left(\frac{a}{b}\right)^n\beta_n\sum_{k=0}^n(-1)^{n-k}{4-k \choose n-k}a^{-k}e_k\big(\sqrt{g^{-1}G}\big)\,.
\eeqn
Interestingly, this has the the same structure as the potential in terms of $\gmn$ and $\fmn$ because it can be written as,
\beqn\label{nepos}
V(g,G)
=2\mu^4\sqrt{g}~\sum_{n=0}^4\beta_n'e_n\big(\sqrt{g^{-1}G}\big)\,,
\eeqn
where the new parameters $\beta_n'$ are,
\beqn\label{newbeta}
\beta_n'=\sum_{m=n}^4(-1)^{m-n}{4-n \choose m-n}\frac{a^{m-n}}{b^m}\beta_m\,.
\eeqn
Note that for the special parameter choice $b/a=\alpha^2 c$, which allows for the existence of proportional backgrounds solutions, the $\beta_n'$ depend on the original $\beta_n$ in a rather complicated way through the solution for $c$. For instance, in the model with $\beta_n=0$ for $n>1$, we get $\beta'_n=0$ for $n>1$ as well as $\beta_0'=\beta_0-\frac{4\beta_1}{\alpha^2c}$ and $\beta_1'=\frac{\beta_1}{a\alpha^2c}$, where $c=-\frac{\beta_0}{6\beta_1}\pm\big(\frac{\beta_0^2}{36\beta_1^2}+\frac{1}{3\alpha^2}\big)^{1/2}$.

The expression for $\fmn$ in terms of $\gmn$ and $G_{\mu\nu}$ also follows directly from the square of~(\ref{sidmg}),
\beqn\label{fmnne}
\fmn=\frac{a^2}{b^2}\left(\gmn-2a^{-1}g_{\mu\rho}{\big(\sqrt{g^{-1}G}\big)^\rho}_\nu+a^{-2}G_{\mu\nu}\right)\,.
\eeqn
Note that this has a structure similar to the expression for $G_{\mu\nu}$ in terms of $\gmn$ and $\fmn$. Of course, it is also possible to obtain a similar expression for $\gmn$. In order to express the entire action in terms of $G_{\mu\nu}$ and $\gmn$, one can now plug~(\ref{fmnne}) into the Einstein-Hilbert term for~$\fmn$ which will give a rather complicated kinetic structure $K(g,G)$ for the fields~$G_{\mu\nu}$ and~$\gmn$. The whole action is of the schematic form,
\beqn\label{actcG}
S(g,G)=m_g^2\int\dd^4x\,\sqrt{g}~R(g)+ m_f^2\int\dd^4x ~K(g,G)-\int\dd^4x~V(g,G)
+S_\mathrm{m}(G,\phi^a)\,.
\eeqn
The potential structure $V(g,G)$ in (\ref{nepos}) is the same as before and the matter coupling of the massless field, $S_\mathrm{m}(G,\phi^a)$ defined in (\ref{matc}), is as simple as in general relativity. Thus, when deciding on the variables for formulating the theory, one has to choose between dealing with the matrix square-root in the matter coupling or with a complicated structure in the kinetic terms. In both formulations it is clearly expected that the theory differs from general relativity at the nonlinear level.

The field $\gmn$ in~(\ref{actcG}) does not have massive fluctuations. In principle, one can now also introduce a nonlinear massive field $M_{\mu\nu}$ and rewrite the action entirely in terms of nonlinear mass eigenstates $G_{\mu\nu}$ and $M_{\mu\nu}$.
In~\cite{Hassan:2012wr} one suggestion for the massive field was $M_{\mu\nu}=F_{\mu\nu}-c\gmn$, with $c$ again being determined in terms of the bimetric parameters through the equation $\Lambda_g=\Lambda_f$. Another possibility is $M_{\mu\nu}=G_{\mu\rho}{(\sqrt{g^{-1}f})^\rho}_\nu-cG_{\mu\nu}$. However, the usefulness of such an approach is questionable since this reformulation will lead to even more complicate structures in the action. Note also that, although classically all formulations are equivalent, it is possible that they will give rise to different quantum theories.

\section{Discussion}\label{discussion}

In this paper we have derived the equations of motion for ghost-free bimetric theory including its coupling to matter through the effective metric $G_{\mu\nu}$. 
As a first application, we studied the proportional background solutions, which correspond to Einstein metrics, along with their perturbation spectrum. 

The quadratic action and linear equations reveal that the fluctuations around proportional backgrounds can be diagonalized into a free massive spin-2 field and a massless spin-2 mode that interacts with matter. At the linear level around backgrounds that admit a definition of mass, the matter-gravity interaction is therefore exactly the same as in general relativity.
On the other hand, the nonlinear theory differs from general relativity because the metric that couples to matter does not possess a standard kinetic term and interacts nontrivially with the massive spin-2 field. 
The existence of proportional backgrounds requires fixing one parameter in the bimetric action with matter coupling, but away from these backgrounds, it is not possible to diagonalize the fluctuation spectrum and identify the massless and massive spin-2 modes. As soon as this single condition on the parameters is imposed the above is true for all choices for the remaining parameters in the action. No further tuning is needed to achieve a decoupling of the massive spin-2 field from the matter sector. Moreover, this parameter choice for which the equations admit proportional background solutions is physically well-motivated because these solutions are degerate with general relativity and hence compatible with observations.

The Boulware-Deser ghost that re-enters bimetric theory above the effective field theory cut-off when the effective metric is coupled to matter is expected to propagate around general solution to the equations of motion derived in this work. However, the nature of the proportional solutions is such that the ghost mode which is present in the full theory does not get excited around these particular backgrounds. The same behaviour was found for perturbations around FRW backgrounds in the massive gravity setup where one metric is nondynamical~\cite{Gumrukcuoglu:2014xba}.

We verified that the effective metric $G_{\mu\nu}$ satisfies the criteria for a nonlinear combination of $\gmn$ and $\fmn$ to be classified as ``massless". This suggests to think of the theory as a nonlinear massless metric~$G_{\mu\nu}$ that interacts with matter and at the same time couples to a nonlinear massive spin-2 field. In this context, an interesting problem to address in bimetric theory with its effective matter coupling is the issue of dark matter. Since the nonlinear massive spin-2 field interacts only with gravity but not directly with matter, it could provide a suitable candidate for the yet unknown dark matter particle. This approach would be different from the one recently taken in~\cite{Aoki:2014cla}, where $\gmn$ and $\fmn$ were coupled to different types of matter and the $\fmn$ matter sector was assumed to account for the dark matter content of the universe.

The fact that, at lowest order in perturbations, matter interacts only with the massless fluctuation of the bimetric sector is expected to have interesting implications for the phenomenology of the model.
In particular, the linear theory around flat space avoids the vDVZ discontinuity~\cite{vanDam:1970vg, Zakharov:1970cc} which leads to unacceptable predictions for observations when a massive spin-2 field is coupled to matter. In linear massive gravity, where matter couples to a massive spin-2 mode, avoiding the vDVZ discontinuity requires the presence of a Vainshtein mechanism~\cite{Vainshtein:1972sx}. 

For the case $b=0$ it is known that cosmological solutions can reproduce the expansion history of the universe~\cite{Volkov:2011an, vonStrauss:2011mq, Comelli:2011zm, Akrami:2012vf}. Moreover, cosmological perturbations for this parameter choice and for the case where the metrics couple to different matter sources, have been studied in~\cite{Comelli:2012db, Berg:2012kn, Fasiello:2013woa, Konnig:2014dna, Comelli:2014bqa, DeFelice:2014nja, Solomon:2014dua, Konnig:2014xva}. In~\cite{Enander:2014xga}, which appeared simultaneously with our work, it is shown that viable cosmological backgrounds can also be obtained for the general case with $b\neq0$.
From our results here it follows that the proportional backgrounds are degenerate with general relativity and therefore give rise to cosmological solutions that can easily be brought to agreement with data by adjusting the bimetric parameters. 
Also the dynamics for the perturbations around these solutions are the same as in general relativity and hence the linear cosmological perturbation theory will be the same. In view of this degeneracy between the linear theory around proportional Einstein backgrounds and linearized general relativity, it will be interesting to study the behavior of perturbations around different backgrounds found in~\cite{Enander:2014xga} that do not admit the notion of mass but still give rise to a viable cosmology.

\vspace{20pt}

\noindent
{\bf Acknowledgments:} The author would like to thank Jonas Enander, Fawad Hassan, Edvard M\"ortsell, Adam Solomon and Mikael von Strauss for numerous helpful discussions.

\appendix

\section{Scalar example}\label{scex}

Here we provide a simple example to illustrate the validity of the identities in~(\ref{rewsmg}) and (\ref{rewsmf}). Instead of the matter action $S_\mathrm{m}$, which depends on the two metrics $\gmn$ and $\fmn$, consider a function $f(x,y)$ of two scalars $x$ and $y$. Suppose this function is of the form,
\beqn
f(x,y)=x+2\sqrt{xy}+y\,,
\eeqn
whose structure is similar to that of the effective metric~(\ref{effg}) appearing in the matter action.
For scalars, the explicit derivatives with respect to $x$ and $y$ are easy to obtain,
\beqn\label{orscv}
\left.\frac{\partial f}{\partial x}\right|_y=1+\frac{\sqrt{y}}{\sqrt{x}}\,,\qquad
\left.\frac{\partial f}{\partial y}\right|_x=1+\frac{\sqrt{x}}{\sqrt{y}}\,.
\eeqn
Alternatively, we can first make the replacement $z\equiv \sqrt{xy}$ such that we have,
\beqn
x(z,y)= \frac{z^2}{y}\,,\qquad
y(z,x)=\frac{z^2}{x}\,.
\eeqn
Plugging these into the function $f(x,y)$ results in,
\beqn
f(x(z,y),y)=\frac{z^2}{y}+2z+y\,,\qquad
f(x,y(z,x))=\frac{z^2}{x}+2z+x\,.
\eeqn
Now consider the variations,
\beqn
\left.\frac{\partial f(x(z,y),y)}{\partial z}\right|_y \left. \frac{\partial z}{\partial x}\right|_y&=&\left(\frac{2z}{y}+2\right)\frac{\sqrt{y}}{2\sqrt{x}}\,,\nn\\
\left.\frac{\partial f(x,y(z,x))}{\partial z}\right|_x \left. \frac{\partial z}{\partial y}\right|_x&=&\left(\frac{2z}{x}+2\right)\frac{\sqrt{x}}{2\sqrt{y}}\,.
\eeqn
Reinserting $z=\sqrt{xy}$ on the right-hand side, we arrive at,
\beqn
\left.\frac{\partial f(x(z,y),y)}{\partial z}\right|_y \left. \frac{\partial z}{\partial x}\right|_y&=&1+\frac{\sqrt{y}}{\sqrt{x}}\,,\nn\\
\left.\frac{\partial f(x,y(z,x))}{\partial z}\right|_x \left. \frac{\partial z}{\partial y}\right|_x&=&1+\frac{\sqrt{x}}{\sqrt{y}}\,.
\eeqn
Comparison with~(\ref{orscv}) shows that we have verified the following identities,
\beqn
\left.\frac{\partial f}{\partial x}\right|_y=\left.\frac{\partial f(x(z,y),y)}{\partial z}\right|_y \left. \frac{\partial z}{\partial x}\right|_y\,,\qquad
\left.\frac{\partial f}{\partial y}\right|_x=\left.\frac{\partial f(x,y(z,x))}{\partial z}\right|_x \left. \frac{\partial z}{\partial y}\right|_x\,.
\eeqn
Although we do not compute the variation of the square-root matrix explicitly in this paper, it should be clear from this simple example that equations~(\ref{rewsmg}) and (\ref{rewsmf}) are indeed valid.

\section{Variations}\label{vari}

Here we provide a few details on the computations of the equations for $\gmn$ and $\fmn$ in section~\ref{dereom}. 
In the following we will vary the expressions,
\beqn
&~~\fmn(F,g)=F_{\mu\rho}g^{\rho\sigma}F_{\sigma\nu}\,,\quad
\gmn(F,f)=F_{\mu\rho}f^{\rho\sigma}F_{\sigma\nu}\,,\nn\\
&~~G_{\mu\nu}(g(F,f),f)=b^2\fmn+2ab\xmn+a^2F_{\mu\rho}f^{\rho\sigma}F_{\sigma\nu}
\,,\nn\\
&G_{\mu\nu}(g,f(F,g))=a^2\gmn+2ab\xmn+b^2F_{\mu\rho}g^{\rho\sigma}F_{\sigma\nu}\,,
\eeqn
with respect to $\xmn$.
The variation of $g^{\alpha\beta}$ with respect to $\xmn$ at fixed $f_{\mu\nu}$ is,
\beqn\label{fvarg}
\left.\frac{\delta g^{\alpha\beta}}{\delta F_{\mu\nu}}\right|_f&=&\left.\frac{\delta g_{\rho\sigma}}{\delta F_{\mu\nu}}\right|_f\frac{\delta g^{\alpha\beta}}{\delta g_{\rho\sigma}}~=~\left.-\frac{1}{2}\frac{\delta g_{\rho\sigma}}{\delta F_{\mu\nu}}\right|_f(g^{\alpha\rho}g^{\beta\sigma}+g^{\alpha\sigma}g^{\beta\rho})\nn\\
&=&-\frac{1}{2}\big(f^{\nu\lambda}F_{\lambda\sigma}\delta^\mu_\rho+f^{\mu\lambda}F_{\lambda\rho}\delta^\nu_\sigma\big)\big(g^{\alpha\rho}g^{\beta\sigma}+g^{\alpha\sigma}g^{\beta\rho}\big)\nn\\
&=&-\frac{1}{2}\big(F^{\nu\beta}g^{\mu\alpha}+F^{\nu\alpha}g^{\mu\beta}+ F^{\mu\beta}g^{\nu\alpha}+F^{\mu\alpha}g^{\nu\beta}\big)\,,
\eeqn
where $F^{\mu\nu}$ with upper indices is the inverse of $\xmn$ and we have used that 
\beqn\label{varfid}
f^{\nu\lambda}F_{\lambda\sigma}g^{\sigma\beta}=f^{\nu\lambda}g_{\lambda\alpha}\big(\sqrt{g^{-1}f}\,\big)^\alpha_{~\sigma}\, g^{\sigma\beta}=\big(\sqrt{f^{-1}g}\,\big)^\nu_{~\sigma} \,g^{\sigma\beta}=F^{\nu\beta}\,.
\eeqn
In turn, $f^{\alpha\beta}$ varied with respect to $\xmn$ with $\gmn$ being fixed reads as,
\beqn\label{fvarx}
\left.\frac{\delta f^{\alpha\beta}}{\delta F_{\mu\nu}}\right|_g&=&\left.\frac{\delta f_{\rho\sigma}}{\delta F_{\mu\nu}}\right|_g\frac{\delta f^{\alpha\beta}}{\delta f_{\rho\sigma}}~=~\left.-\frac{1}{2}\frac{\delta f_{\rho\sigma}}{\delta F_{\mu\nu}}\right|_g(f^{\alpha\rho}f^{\beta\sigma}+f^{\alpha\sigma}f^{\beta\rho})\nn\\
&=&-\frac{1}{2}\big(g^{\nu\lambda}F_{\lambda\sigma}\delta^\mu_\rho+g^{\mu\lambda}F_{\lambda\rho}\delta^\nu_\sigma\big)\big(f^{\alpha\rho}f^{\beta\sigma}+f^{\alpha\sigma}f^{\beta\rho}\big)\nn\\
&=&-\frac{1}{2}\big(F^{\nu\beta}f^{\mu\alpha}+F^{\nu\alpha}f^{\mu\beta}+ F^{\mu\beta}f^{\nu\alpha}+F^{\mu\alpha}f^{\nu\beta}\big)\,,
\eeqn
where again we have made use of (\ref{varfid}). As a result, the equations of motions~(\ref{neweqgf}) assume the form, 
\beqn\label{gfeqwj}
\left.\frac{\delta g^{\rho\sigma}}{\delta\xmn}\right|_{f}\left.\frac{\delta S}{\delta g^{\rho\sigma}}\right|_{f}&=&
(F^{\rho\mu}g^{\sigma\nu}+F^{\rho\nu}g^{\sigma\mu})
\left.\frac{\delta S}{\delta g^{\rho\sigma}}\right|_{f}~=~0\,,\nn\\
\left.\frac{\delta f^{\rho\sigma}}{\delta\xmn}\right|_{g}\left.\frac{\delta S}{\delta f^{\rho\sigma}}\right|_{g}&=&
(F^{\rho\mu}f^{\sigma\nu}+F^{\rho\nu}f^{\sigma\mu})
\left.\frac{\delta S}{\delta f^{\rho\sigma}}\right|_{g}~=~0\,.
\eeqn
Varying $G^{\alpha\beta}$ with respect to $\xmn$ keeping $\fmn$ fixed gives,
\beqn\label{Gvarx2}
&~&\left.\frac{\delta G^{\alpha\beta}(g(F,f),f)}{\delta F_{\mu\nu}}\right|_f=\left.\frac{\delta G_{\rho\sigma}}{\delta F_{\mu\nu}}\right|_f\frac{\delta G^{\alpha\beta}}{\delta G_{\rho\sigma}}~=~\left.-\frac{1}{2}\frac{\delta G_{\rho\sigma}}{\delta F_{\mu\nu}}\right|_f(G^{\alpha\rho}G^{\beta\sigma}+G^{\alpha\sigma}G^{\beta\rho})\nn\\
&~&\hspace{50pt}=-\frac{1}{2}\big(2ab\,\delta^\mu_\rho\delta^\nu_\sigma+a^2f^{\mu\lambda} F_{\lambda\sigma}\delta^\nu_\rho+a^2f^{\nu\lambda} F_{\lambda\sigma}\delta^\mu_\rho\big)\big(G^{\alpha\rho}G^{\beta\sigma}+G^{\alpha\sigma}G^{\beta\rho}\big)\,.
\eeqn
Finally, the variation of $G^{\alpha\beta}$ with respect to $\xmn$ at fixed $\gmn$ is,
\beqn\label{Gvarx}
&~&\left.\frac{\delta G^{\alpha\beta}(g,f(F,g))}{\delta F_{\mu\nu}}\right|_g=\left.\frac{\delta G_{\rho\sigma}}{\delta F_{\mu\nu}}\right|_g\frac{\delta G^{\alpha\beta}}{\delta G_{\rho\sigma}}~=~\left.-\frac{1}{2}\frac{\delta G_{\rho\sigma}}{\delta F_{\mu\nu}}\right|_g(G^{\alpha\rho}G^{\beta\sigma}+G^{\alpha\sigma}G^{\beta\rho})\nn\\
&~&\hspace{50pt}=-\frac{1}{2}\big(2ab\,\delta^\mu_\rho\delta^\nu_\sigma+b^2g^{\mu\lambda} F_{\lambda\sigma}\delta^\nu_\rho+b^2g^{\nu\lambda} F_{\lambda\sigma}\delta^\mu_\rho\big)\big(G^{\alpha\rho}G^{\beta\sigma}+G^{\alpha\sigma}G^{\beta\rho}\big)\,.
\eeqn



\end{document}